\documentclass[prl,twocolumn,showpacs,preprintnumbers,amsmath,amssymb, superscriptaddress]{revtex4}

\usepackage{graphicx}
\usepackage{dcolumn}
\usepackage{bm}

\begin{document}


\title{Post-selective two-photon interference from a continuous non-classical stream of photons emitted by a quantum dot}

\author{R. B. Patel}\email{rp349@cam.ac.uk}
\affiliation{Toshiba Research Europe Limited, Cambridge Research Laboratory, 208 Cambridge Science Park, Milton Road, Cambridge, CB4 0GZ, U. K.}
\affiliation{Cavendish Laboratory, Cambridge University, JJ Thomson Avenue, Cambridge, CB3 0HE, U. K.}
\author{A. J. Bennett}
\affiliation{Toshiba Research Europe Limited, Cambridge Research Laboratory, 208 Cambridge Science Park, Milton Road, Cambridge, CB4 0GZ, U. K.}
\author{K. Cooper}
\affiliation{Cavendish Laboratory, Cambridge University, JJ Thomson Avenue, Cambridge, CB3 0HE, U. K.}
\author{P. Atkinson}
\altaffiliation[Now at: ]{Max-Planck-Institut f\"{u}r Festk\"{o}rperforschung, Heisenbergstr.1, D-70569 Stuttgart, Germany.\\}
\author{C. A. Nicoll}
\affiliation{Cavendish Laboratory, Cambridge University, JJ Thomson Avenue, Cambridge, CB3 0HE, U. K.}
\author{D. A. Ritchie}
\affiliation{Cavendish Laboratory, Cambridge University, JJ Thomson Avenue, Cambridge, CB3 0HE, U. K.}
\author{A. J. Shields}
\affiliation{Toshiba Research Europe Limited, Cambridge Research Laboratory, 208 Cambridge Science Park, Milton Road, Cambridge, CB4 0GZ, U. K.}

\date{\today}

\begin{abstract}
We report an electrically driven semiconductor single photon source capable of emitting photons with a coherence time of up to 400 ps under fixed bias.  It is shown that increasing the injection current causes the coherence time to reduce and this effect is well explained by the fast modulation of a fluctuating environment.  Hong-Ou-Mandel type two-photon interference using a Mach-Zehnder interferometer is demonstrated using this source to test the indistinguishability of individual photons by post-selecting events where two photons collide at a beamsplitter.  Finally, we consider how improvements in our detection system can be used to achieve a higher interference visibility.
\end{abstract}

\maketitle
The interference of two photons at a beamsplitter has been the focal point of research in linear optical quantum information processing in recent years. This effect, often referred to as Hong-Ou-Mandel \cite{hom87} (HOM) interference, arises when the probability amplitudes of two-photon states destructively interfere.  Provided the single-photon wavefunctions overlap perfectly at the beamsplitter and that they are indistinguishable in the spatial, temporal, spectral and polarization degrees of freedom they should always exit the beamsplitter along the same port.  Two-photon interference was first observed using sources of parametric down converted photons \cite{hom87,shih88} but the proposal of Knill, Laflamme, and Milburn \cite{klm01}, in particular, has spurred much research effort into developing on-demand sources of single photons.  Single photon sources such as single molecules \cite{kiraz05}, trapped ions \cite{maunz07}, atoms \cite{beugnon06}, and semiconductor quantum dots \cite{santori02} have been used to demonstrate two-photon interference.  These sources have all relied upon quasi-resonant laser excitation to generate identical single photon states.

Incoherent pumping of a two level emitter can lead to processes which induce homogeneous broadening of the quantum state and a reduction in coherence.  In regard to two-photon interference, dephasing destroys the indistinguishability of the individual photons.  For exciton recombination in semiconductor quantum dots this is especially true as the quantum dot is able to interact with phonons and localized carriers in the surrounding semiconductor. Dephasing of quantum dots as a function of temperature \cite{borri01,kammerer02,bayer02,berthelot06} and laser excitation density \cite{kammerer02PRB,gotoh04,berthelot06,favero07} has been studied extensively in recent years.  In the latter case dephasing can be attributed to Coulombic interactions between carriers inside or outside the quantum dot \cite{uskov01,gotoh04}.

In this paper we measure the variation in coherence time as a function of DC current injection in a microcavity LED.  In contrast to pulsed excitation schemes, this allows a  HOM-type two-photon interference experiment to be carried out without having to match delays in our interferometer and with the benefit of higher count rates.

\begin{figure}
\includegraphics{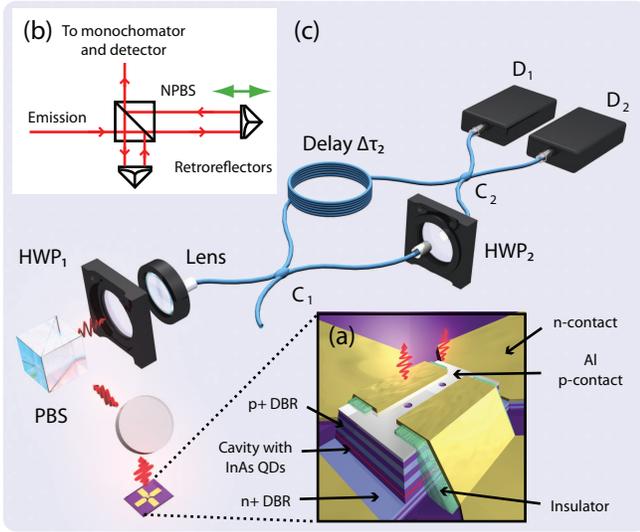}
\caption{\label{fig:Setup} (Colour Online).  (a) An illustration of our device. (b) Arrangement of the Michelson interferometer used to measure the coherence time.  The emission is divided and recombined at different points on a non-polarizing beam splitter (NPBS).  (c) Mach-Zehnder interferometer for measuring two-photon interference.  Horizontally polarized photons are selected using a polarizing beamsplitter (PBS) and a half-wave plate (HWP$_1$) aligns the polarization to the axis of the polarization-maintaining single-mode fiber.  A monochromator (omitted for clarity) located between HWP$_1$ and the lens is used to filter the emission.  The emission is then coupled into the fiber and split at the first coupler C$_1$.  The two arms can be made distinguishable or indistinguishable by rotating HWP$_2$. For indistinguishable photons interference occurs at the final coupler C$_2$ resulting in a suppression of coincident counts at the two avalanche photodiode detectors D$_1$ and D$_2$.}
\end{figure}

Our sample, shown in Fig. \ref{fig:Setup}(a), is a microcavity p-i-n diode\cite{yuan02,bennett05} consisting of two (twelve) GaAs/Al$_{0.98}$Ga$_{0.02}$As layers forming a distributed Bragg reflector above (below) a $\lambda$ cavity,  with a layer of InGaAs/GaAs quantum dots at its center.  An aluminium mask on top of the mesa ($40\times40 \textrm{ }\mu\textrm{m}$ area) with $\sim2\textrm{ }\mu\textrm{m}$ diameter apertures acts as a p-contact and allows single quantum dots to be isolated.  Due to the large modal volume and low $Q$-factor of the cavity there is no measurable Purcell effect and the cavity merely serves to enhance the collection efficiency.  The sample is cooled to 4 K in a continuous flow cryostat and the emission from the dot is collected using an objective lens.  A polarizing beamsplitter (PBS) enables horizontally polarized photons to be selected.

\begin{figure}
\includegraphics{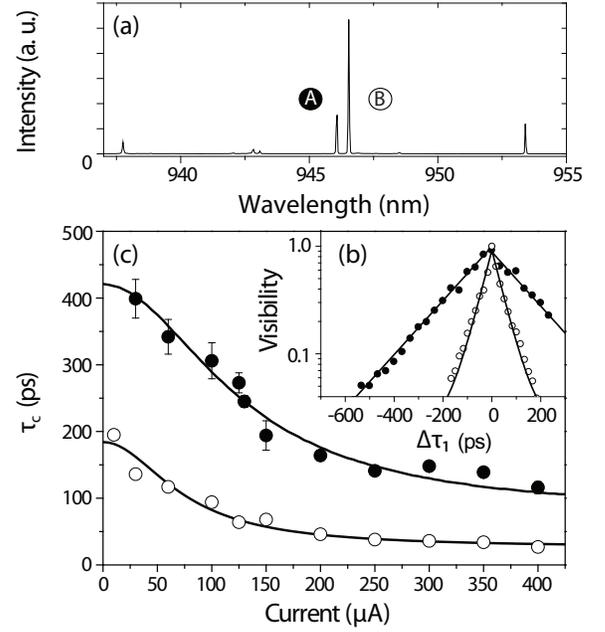}
\caption{\label{fig:cohdata} (a) Electroluminescence spectra showing the two charged exciton states.  (b) Typical visibility plots taken with an injection current of 200 $\mu$A. Solid circles correspond to line A emitting at 946.3 nm and empty circles to line B emitting at 946.8 nm. (c) Measurements of the coherence time $\tau_c$ as a function of current.  The black lines are the theoretical fits.  }
\end{figure}

In Fig. \ref{fig:Setup}(b) the electroluminescence from the quantum dot is passed through a Michelson interferometer where Fourier transform spectroscopy \cite{kammerer02} is carried out to deduce the coherence time $\tau_{c}$ of the emitting state.  The electroluminescence spectra in Fig. \ref{fig:cohdata}(a) show two bright lines, line A emitting at 946.3 nm and line B emitting at 946.8 nm.  The absence of fine-structure splitting suggests that they are both charged exciton states. Adjusting the position of a lens placed in front of the CCD detector allows the relative intensities of the two lines to be changed which also suggests that they correspond to two spatially separate dots.

Measurements of the single photon interference visibility vary as $\sim\exp(-|\Delta{\tau_{1}}|/\tau_{c})$ where $\Delta{\tau_{1}}$ is the time delay between the arms of the interferometer.  Fig. \ref{fig:cohdata}(b) shows typical plots of the visibility we observe for the range of injection currents studied.  In Fig. \ref{fig:cohdata}(c) we present the variation in coherence time of the two states as a function of injection current.  For line A and a current of 30 $\mu$A we measure a coherence time of 400 ps which, to the best of our knowledge, is the largest value reported for a InGaAs/GaAs quantum dot embedded in a microcavity, in any excitation scheme.  In both cases, as the current is increased a reduction in coherence is observed.  We attribute this dephasing to charge fluctuations in the vicinity of the quantum dot.  Fluctuations of this type may arise from impurities or defects in the wetting layer which result in a variation of the emission wavelength over time via the quantum confined Stark effect.  In what follows, we adopt the approach of \citet{favero07} and model the variation in coherence time.  In this regime a Stark shift $\Delta$ is produced by $N$ individual traps which randomize the emission energy of the state over a range given by the modulation amplitude\cite{berthelot06} $\Sigma = 2\Sigma_{s}/(\sqrt{\tau_{\uparrow}/\tau_{\downarrow}}+\sqrt{\tau_{\downarrow}/\tau_{\uparrow}})$ where $\Sigma_{s} = \sqrt{N}\Delta/2$ is the saturation value.  This process occurs on a characteristic timescale $\tau_{f}$ given by $1/\tau_{f} = 1/\tau_{\uparrow} + 1/\tau_{\downarrow}$. The rate of capture $1/\tau_{\downarrow}$ and escape $1/\tau_{\uparrow}$ are given by the rate equations
\begin{equation}\label{eq:taucapture}
    \frac{1}{\tau_{\downarrow}} = \frac{1}{\tau_{2}}(1 + n_{2})
\end{equation}
\begin{equation}\label{eq:tauescape}
    \frac{1}{\tau_{\uparrow}} = \frac{1}{\tau_{1}}n_{1} + \frac{1}{\tau_{3}}\left(\frac{I^{\beta}}{I^{\beta} + I^{\beta}_{0}}\right)
\end{equation}
Terms $n_{i}$ are the Bose-Einstein occupation factors given by $1/(\exp(E_{i}/k_{B}T)-1)$.  Terms involving a subscript 1 (2) pertain to acoustic (optical) phonon emission or absorption which can lead to carrier capture or escape respectively.  The characteristic timescale for Auger processes is represented by $\tau_{3}$ and $I_{0}$ is the current at which the Auger process saturates.  The solid lines in Fig. \ref{fig:cohdata}(c) show the theoretical variation in coherence time $\tau_{c} = \hbar^{2}/\Sigma^{2}\tau_{f}$ as a function of current using similar parameters as Ref. \cite{favero07} such that $\tau_{1} = 200 \textrm{ ps}$, $\tau_{2} = 5 \textrm{ ps}$, $E_{1} = 1 \textrm{ meV}$, $E_{2} = 30 \textrm{ meV}$, and $\beta = 2$ but with the fitting parameters $\tau_{3} = 750 \textrm{ ps}$, $I_{0} = 300 \textrm{ }\mu\textrm{A}$,  and $\Sigma_{s} = 188 \textrm{ }\mu\textrm{eV}$ for line A and $\tau_{3} = 550 \textrm{ ps}$, $I_{0} = 200 \textrm{ }\mu\textrm{A}$, and $\Sigma_{s} = 285 \textrm{ }\mu\textrm{eV}$ for line B.  In Ref. \cite{berthelot06} a Gaussian component was observed at high excitation density owing to inhomogeneous broadening of the state and the ratio $\Sigma\tau_f/\hbar \geq 1$.  In our excitation scheme, even with a large injection current of $200 \textrm{ }\mu\textrm{A}$, this ratio is calculated to be 0.01 and 0.03 for lines A and B respectively so that a Lorentzian line shape is observed \cite{favero07}. It is evident from these plots that the model of spectral diffusion of the transition line, due to the asymmetric efficiencies of the capture and escape processes, describes our data well.

\begin{figure*}
\includegraphics{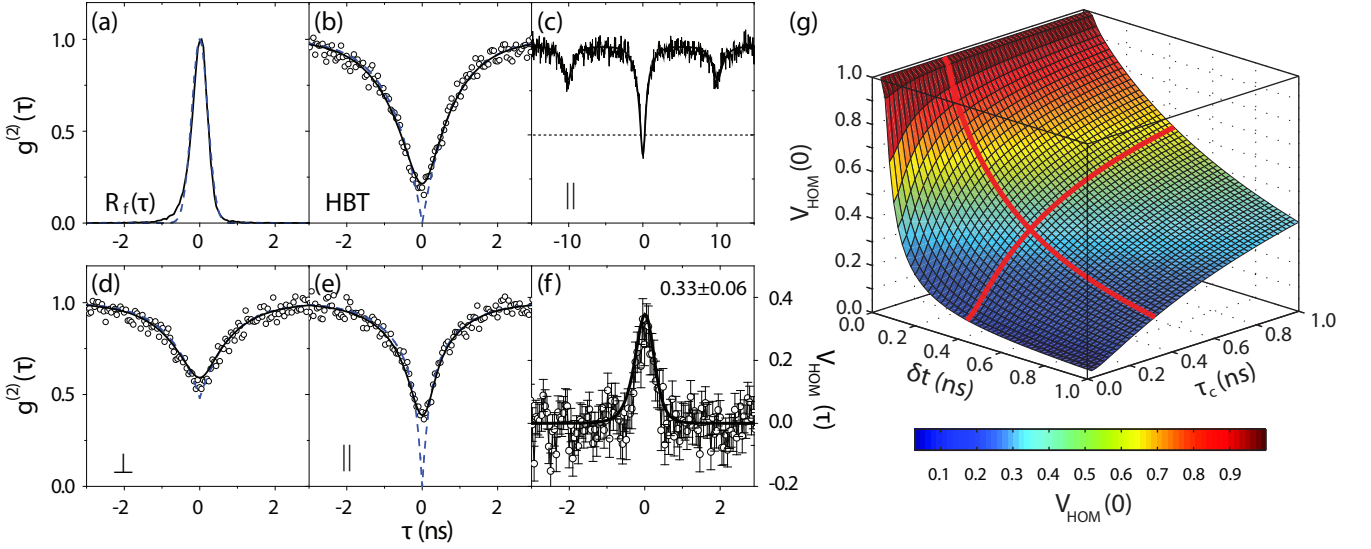}
\caption{\label{fig:HOM} (Colour Online). Hanbury-Brown and Twiss and two-photon interference results. (a) The measured system response (solid line) and its Gaussian approximation with FWHM $\delta{t} = 428 \textrm{ }ps$ (dashed line).  In (b) to (e) dashed lines show ideal curves without detection system limitation and bold lines model the effect of finite system response.  (b) $g^{(2)}(\tau)$.  (c) $g^{(2)}_{\parallel}(\tau)$, dotted line indicates the classical limit.  (d) $g^{(2)}_{\perp}(\tau)$.  (e) Detailed plot of (c).  (f) Two-photon interference visibility.  (g)  Variation in $V_{\textrm{HOM}}$ with system resolution $\delta{t}$ and coherence time $\tau_c$.  The bold lines correspond to our experimental conditions.}
\end{figure*}

The intensity of line A increased and saturated at $\approx 200 \textrm{ }\mu\textrm{A}$.  We therefore chose to operate the device with a current of 100 $\mu\textrm{A}$ at which point the intensity was approximately half the saturated value.  This allowed for two-photon interference measurements to be carried out with a relatively long coherence time of $\sim 325 \textrm{ }\textrm{ps}$, corresponding to a line width of $4 \textrm{ }\mu\textrm{eV}$.  We now present the main result of this paper.

Two-photon interference measurements were carried out under DC operation using a fiber-coupled Mach-Zehnder interferometer depicted in Fig. \ref{fig:Setup}(c).  A polarizing beamsplitter (PBS) was used to select horizontally polarized photons and half-wave plate HWP$_{1}$ was adjusted to align the polarization of the emission to the birefringence axis of the polarization-maintaining single-mode fiber.  This ensures that the polarization state of the photons is conserved in the fiber. The emission was then filtered using a monochromator, set to select a spectral width of $88 \textrm{ }\mu\textrm{eV}$, (not shown) and coupled into the interferometer.  A fiber coupler C$_{1}$ split the stream of photons along two paths, one of which contained a delay $\Delta{\tau_{2}} = 10$ ns and the other a second half-wave plate (HWP$_{2}$). HWP$_{2}$ was rotated to make the polarization of photons in each arm mutually parallel or orthogonal thereby making the paths indistinguishable or distinguishable, respectively.  When the paths were indistinguishable, two photons traveling along each arm and arriving at the second $2\times2$ fiber coupler C$_{2}$, within the single-photon coherence length, interfered destructively leading to a suppression in coincident counts at the two avalanche photodiode detectors D$_{1}$ and D$_{2}$.  The delay $\Delta{\tau_{2}}$, which was much greater than the coherence length of the individual photons, ensured that only fourth-order interference effects occur at C$_{2}$.  Using single-mode fiber enabled the spatial modes to be easily matched at C$_{2}$.

The quantities of interest are the second-order coherence functions $g^{(2)}(\tau)$ as measured using a Hanbury-Brown and Twiss (HBT) arrangement, and $g^{(2)}_{\bot}(\tau)$ and $g^{(2)}_{\|}(\tau)$ which describe correlations at the two detectors when the photons have orthogonal and parallel polarizations respectively.  Here $\tau$ is the delay between detections.  In the limit well below saturation of the state these can be expressed as
\begin{eqnarray}
    g^{(2)}(\tau) &=& 1 - e^{-|\tau|/\tau_{r}}\label{eq:g2}\\
    g^{(2)}_{\bot}(\tau) &=& 4\left(T_{1}^{2} + R_{1}^{2}\right)R_{2}T_{2}g^{(2)}(\tau)\nonumber\\
    &+& 4R_{1}T_{1}\Big(T_{2}^{2}g^{(2)}(\tau-\Delta{\tau_2})\nonumber\\
    &+& R_{2}^{2}g^{(2)}(\tau+\Delta{\tau_2})\Big)\label{eq:g2perp}\\
    g^{(2)}_{\|}(\tau) &=& 4\left(T_{1}^{2} + R_{1}^{2}\right)R_{2}T_{2}g^{(2)}(\tau)\nonumber\\ &+& 4R_{1}T_{1}\left(T_{2}^{2}g^{(2)}(\tau-\Delta{\tau_2})+ R_{2}^{2}g^{(2)}(\tau+\Delta{\tau_2})\right)\nonumber\\&\times& \left(1-Ve^{-2|\tau|/\tau_{c}}\right)\label{eq:g2para}
\end{eqnarray}
\noindent where $R$ and $T$ represent the reflection and transmission intensity coefficients of the two fiber couplers, $\tau_{r}$ is the radiative lifetime, and $V$ is a function which is dependent on the overlap of the wavefunctions at C$_{2}$.  For orthogonally polarized photons, classical correlations occur at the two detectors and for a perfect single photon source coincident counts are expected to occur fifty percent of the time.  This allows the two-photon interference visibility to be defined as  $V_{\textrm{HOM}}(\tau) = \left(g^{(2)}_{\perp}(\tau)-g^{(2)}_{\parallel}(\tau)\right)/g^{(2)}_{\perp}(\tau)$.

The response of our system $R_f(\tau)$, shown as the solid line in  Fig. \ref{fig:HOM}(a), was measured by taking a HBT correlation for photons emitted by a mode-locked Ti-Sapphire laser tuned to 940 nm.  All HBT measurements were carried out using a separate non-polarizing $2\times2$ coupler (not shown). The function $R_f(\tau)$ is limited by the response time of our APDs.  The slight asymmetry is due to their unequal individual responses to a short impulse.  An additional HBT correlation shown in Fig. \ref{fig:HOM}(b) (open circles) was taken to determine $g^{(2)}(\tau)$ for our source.  Equation \ref{eq:g2} is also plotted for comparison (dashed line).  We find that by plotting $g^{(2)}(\tau) \otimes R_f(\tau)$ with $g^{(2)}(0)=0$ and $\tau_{r} = 800\textrm{ ps}$ we are able to produce a good fit to the data (bold line).  It is therefore reasonable to assume that for our source $g^{(2)}(0) \approx 0$ and that the radiative lifetime of the state is 800 ps.

In Fig. \ref{fig:HOM}(c) we present a two-photon interference correlation taken for photons with parallel polarizations.  We observe a dip at zero delay below the classical limit, indicated by the dotted line, and two dips down to 0.75 at $\pm$ 10 ns due to the delay in our interferometer.  From Eq. (\ref{eq:g2para}), equality of these two dips suggests that the final coupler is balanced and $R_2=T_2=0.5$.  The bold line shows $g^{(2)}_{\|}(\tau) \otimes R_f(\tau)$ with $\tau_{c} = 325\textrm{ ps}$ as measured in Fig. \ref{fig:cohdata}(c).  Figure \ref{fig:HOM}(d) and (e) show detailed plots of the measured second-order coherence functions around zero delay (open circles) for orthogonal and parallel photons respectively.  Again Eq. (\ref{eq:g2perp}) and (\ref{eq:g2para}) are also plotted for comparison (dashed lines) along with $g^{(2)}_{\bot}(\tau) \otimes R_f(\tau)$ and $g^{(2)}_{\|}(\tau) \otimes R_f(\tau)$ (bold lines).  We see that in the absence of any fitting parameters our fits are in excellent agreement with the experimental data.  It is evident that for photons of parallel polarization the suppression at zero delay is limited by the detector response.  The observed two-photon interference visibility of $0.33 \pm 0.06$ (see Fig. \ref{fig:HOM}(f)) is consistent with the assumption that interference is entirely limited by the resolution of the detection system and that there is 100$\%$ overlap of the photon wavefunctions.

We now consider whether it is possible to use this method to post-select a higher visibility of interference.  Using Eq. (\ref{eq:g2perp}) and (\ref{eq:g2para}) and a Gaussian system response (Fig. \ref{fig:HOM}(a)) we are able to estimate the visibility of interference under different experimental conditions.  In contrast to pulsed two-photon interference \cite{santori02}, the figure of merit in this case is not $2\tau_{r}/\tau_{c}$ but rather $2\delta{t}/\tau_{c}$, which should be minimized in order to observe a high interference visibility.  The bold lines in Fig. \ref{fig:HOM}(g) indicate the range of visibilities for $\delta{t} = 428$ ps and $\tau_{c} = 325$ ps corresponding to our experiment.  From Fig. \ref{fig:cohdata}(b) we infer that reducing the current to 30 $\mu$A and using the same detection system would result in increasing $\tau_{c}$ to 400 ps and the visibility to $\sim$45\%.  On the other hand, reducing the system timing resolution to $\sim$100 ps should be sufficient to observe a visibility greater than 70\% which could be achieved using superconducting single-photon detectors \cite{gol'tsman05}.

In conclusion we have shown that dephasing processes affecting an electrically driven quantum dot are well described by the fast modulation of a fluctuating charge environment.  By using a low current it is possible to generate photons with a coherence time of several hundred picoseconds.  With an appropriate detection system it would be possible to observe high visibility HOM type two-photon interference, using an electrically driven single photon source, suitable for applications such as tests against local realism \cite{aspect82} and entanglement swapping \cite{halder07}.

This work was partially funded by the EU-projects QAP and SANDiE.  One of the authors (R. B. P.) would also like to thank EPSRC and TREL for funding.



\end{document}